\documentclass[twocolumn,pre,aps,showpacs,amsmath,amssymb]{revtex4-1}
\usepackage{graphicx}
\usepackage{bm}
\usepackage{epstopdf}
\begin{document}

\title{Nonequilibrium thermodynamics of self-supervised learning}
\author{Domingos S. P. Salazar}
\email[]{salazar.domingos@gmail.com}
\affiliation{Unidade de Educa\c{c}\~{a}o a Dist\^{a}ncia e Tecnologia, Universidade Federal Rural de Pernambuco, Recife, Pernambuco 52171-900 Brazil}


\begin{abstract}
Self-supervised learning (SSL) of energy based models has an intuitive relation to equilibrium thermodynamics because the softmax layer, mapping energies to probabilities, is a Gibbs distribution. However, in what way SSL is a thermodynamic process? We show that some SSL paradigms behave as a thermodynamic composite system formed by representations and self-labels in contact with a nonequilibrium reservoir. Moreover, this system is subjected to usual thermodynamic cycles, such as adiabatic expansion and isochoric heating, resulting in a generalized Gibbs ensemble (GGE). In this picture, we show that learning is seen as a demon that operates in cycles using feedback measurements to extract negative work from the system. As applications, we examine some SSL algorithms using this idea.
\end{abstract}
\pacs{07.05.Mh, 05.70.Ln}
\maketitle

\section{Introduction}
\label{intro}
Representation learning is commonly seen as adjusting weights of controllable parameters of a model so it better represents data. In some theoretical analysis and engineering applications \cite{DeepReview,Goodfellow2016,RMPReview2019,Lecun1990,HintonScience2006,Alexnet}, learning is supervised, meaning that true labels teach the system how to fix current mistakes. However, in nature and in the next frontier of large scale engineering applications, learning is most likely to happen in the absence of labels \cite{Hinton2002,HintonDBN,Lecun2007,HintonTrends2007}, a situation known as self-supervised learning (SSL). In this case, what is to be learned? Although one cannot check if a data point belongs to a certain category as in the supervised case, a SSL model can still learn a good \textit{representation} of data. One could think of it as a low dimensional vector that encodes what is actually relevant in representing data.

In situations where annotated (labeled) data is expensive or scarce, SSL became the source of recent artificial intelligence breakthroughs \cite{Deepcluster,Moco,Simclr,Byol,Swav,Seer,Asano2020,Barlowtwins}. In part because of the huge size of unlabelled data sets of computer vision \cite{Seer} and natural language processing \cite{Bert}, but also due to highly polished sensory modules \cite{Resnet,Regnet} of deep learning architectures.

Also recently, there was a rising interest in bringing machine learning (ML) and physics closer \cite{Wu2019,Sharir2020,Barbier2019,Carleo2017,Nguyen2017,Decelle2017}, particularly thermodynamics \cite{SeifertPRL2017,Seifert2017b,Salazar2017}, motivated by concepts from stochastic thermodynamics and nonequilibrium, such as irreversible work, entropy production and fluctuation theorems \cite{Jar1997,Crooks1999,Jar2004,Seifert2008,Sekimoto2010,Seifert2012}. Despite that, the pace of SSL advances are not remotely matched by their thermodynamics interpretations. For instance, some recent results in SSL successfully applied self-labelling approaches \cite{Asano2020,Deepcluster,Dosovitskiy2016,Zhirong2018,Swav,Seer} with remarkable performance, but how this idea translates into thermodynamics is not clear.

\begin{figure}[htp]
\includegraphics[width=3.3 in]{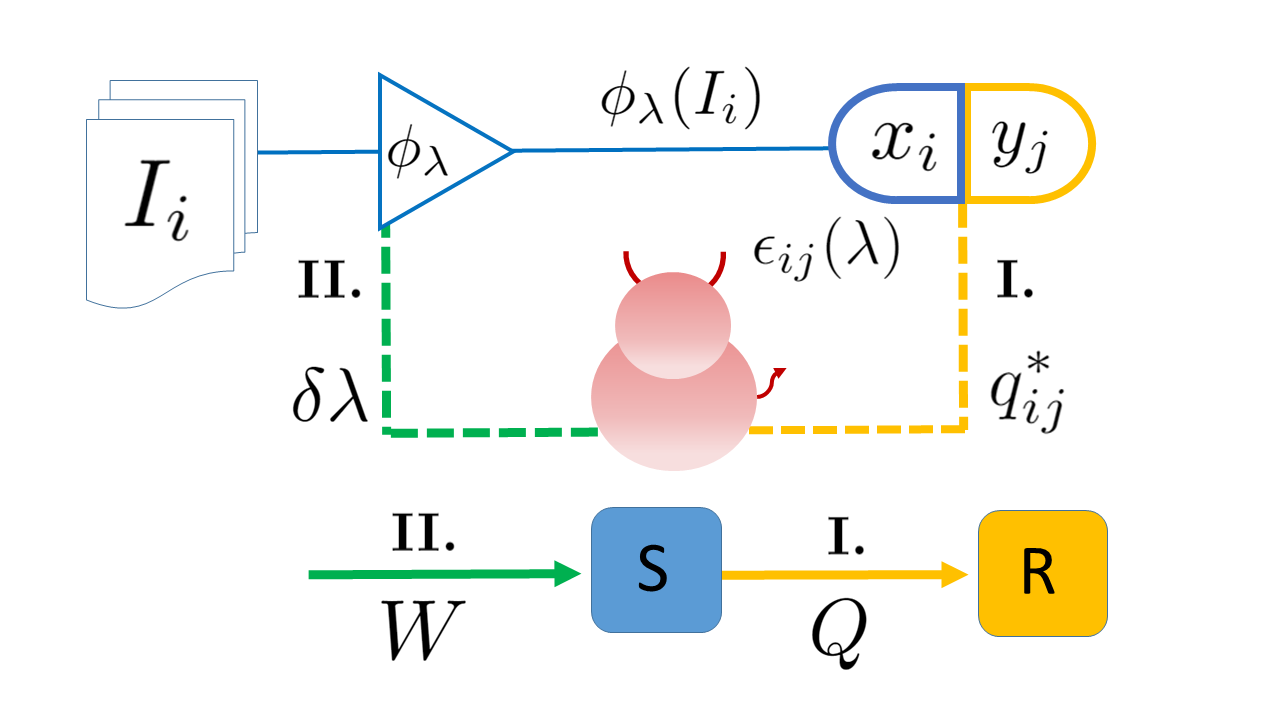}
\caption{(Color online) SSL as a demon: $N$ data points $I_i$ are mapped into representations $x_i$ by a function $\phi_\lambda$. A composite system $S$ is formed by the representations and a set of $K$  self-labels $y_j$, where each state $(i,j)$ has some energy $\epsilon_{ij}(\lambda)$ used to define stochastic heat and work. I. Heating ($W=0$) takes the system to a GGE, $q^*_{ij}$ (yellow). II. Then, in an adiabatic process ($Q=0$), the demon uses current knowledge of $\epsilon_{ij}(\lambda)$ and picks a small $\delta \lambda$ so it performs negative (irreversible) work in the system (green), $W_{irr}<0$, repeating the cycle.}
\label{Demon.png}
\end{figure}

In this paper, we show that a class of algorithms \cite{Swav,Seer,Deepcluster} that uses a form of optimal transport for self-labelling are actually working as a thermodynamic Otto half-cycle. This is achieved by defining a composite system formed by the representations $i$ and self-labels $j$ (FIG. 1), described as an ensemble $q_{ij}$. Then, using the maximum entropy principle (MaxEnt), the online self-labelling is interpreted as a heating process that leads the system to a generalized Gibbs ensemble (GGE) with general constraints. Adding some particular constraints reduces the GGE to a solution of an optimal transport problem which is solved using smooth version of the Sinkhorn-Knopp algorithm \cite{Asano2020,Cuturi2013}. Finally, we show that the subsequent learning step is a demon that extracts negative work in a adiabatic process using information from the system.

In this analogy, the GGE plays a central role defining self-labelling strategies. In principle, one could craft new SSL algorithms by choosing suitable constraints for the GGE and a composition of thermodynamic processes.

The paper is organized as follows. Section II introduces the formalism for SSL using thermodynamics and some processes used as building blocks of the algorithms. Section III casts the learning problem as a demon operating in cycles. Section IV applies the ideas to classes of algorithms, specially the Sinkhorn-Knopp class, followed by Section V with conclusions.

\section{Formalism}
We consider $N$ data points $\{I_1,...,I_N\}$ with $I_i \in \mathbb{R}^D$ (for instance, images) and $K$ classes $\{y_1,...,y_K\}$ (self-labels) such that each pair $(I_i, y_j)$ defines a possible state of a composite system. Let $\phi_\lambda$ be a function that maps data points $I_i$ to representations, $x_i = \phi_{\lambda}(I_i) \in \mathbb{R}^R$. The subscript $\lambda$ indicates that the function $\phi_\lambda$ depends on a set of weights, $\lambda \in \mathbb{R}^W$. The function $\phi_\lambda$ is the output of a deep neural network, where typically $R\ll D$, although the properties derived in this formalism have general purpose. Define a scalar energy $E_{\lambda}(x,y)$, for which its observable values over the allowed states take the form of a matrix,
\begin{equation}
\label{maths}
\epsilon_{ij}(\lambda):=E_{\lambda}(\phi_\lambda(I_i),y_j),
\end{equation}
that depends explicitly on the state $(I_i,y_j)$ and the weights $\lambda$. Now we introduce an ensemble $q_{ij}$ such that for each pair $(i,j)$, we assign a probability $0\leq q_{ij}\leq 1$ such that $\sum_{ij} q_{ij}=1$. The internal energy is defined as the average stochastic energy over the ensemble,
\begin{equation}
\label{internalenergy}
\langle E_{\lambda}(x,y) \rangle = \sum_{ij} \epsilon_{ij}(\lambda)q_{ij}. 
\end{equation}
For a transformation $q_{ij}\rightarrow q_{ij}'$ and $\lambda \rightarrow \lambda'$, it results in the following definitions of work and heat within the framework of stochastic thermodynamics \cite{Seifert2012,Sekimoto2010}
\begin{eqnarray} 
\label{work}
W:=\sum_{ij}q_{ij}(\epsilon_{ij}(\lambda')-\epsilon_{ij}(\lambda)),
\\
\label{heat}
Q:=\sum_{ij}(q_{ij}-q_{ij}')\epsilon_{ij}(\lambda'),
\end{eqnarray}
where the First Law of Thermodynamics holds, $\langle \Delta E_{\lambda}(x,y) \rangle = W - Q$.
Note that, in this formalism, supervised learning is a special ensemble, where $q_{ij}=\delta_{jk}/N$, for $y_k$ the true label of $I_i$. In the absence of true labels, SSL creates self-labels with some uncertainty modeled by the ensemble $q_{ij}$. As we aim to understand thermodynamic cycles, we consider a single minibatch of $N$ data points for simplicity. However, considering several random mini-batches is actually feasible in stochastic thermodynamics and it could be interpreted as randomness in initial conditions (due to, for instance, a physical system of small size).

Note that the definitions (\ref{work},\ref{heat}) are typical from stochastic thermodynamics and the SSL mechanism was not presented yet. Actually, devising a mechanism to find an ensemble $q_{ij}$ iteratively is a merit of the underlying SSL algorithm. Our contribution is to show that, in some cases, finding $q_{ij}$ is analogous to subjecting the composite system to a thermodynamic process of isochoric heating. In this case, the analogy between SSL and thermodynamics goes beyond simple definitions of stochastic heat and work, entering in the core of maximum entropy ensembles.

\subsection{Heating process} Isochoric heating changes the ensemble $q_{ij}\rightarrow q_{ij}'$ (or occupations) without changing the weights. Therefore, work in (\ref{work}) is zero, $W=0$, and heat is given by (\ref{heat}) for $\lambda'=\lambda$,
\begin{equation}
    \label{isoheat}
Q=\sum_{ij}(q_{ij}-q_{ij}')\epsilon_{ij}(\lambda),
\end{equation}
Physically, one could think of a gas with fixed volume (ie, weigths) exchanging heat with a thermal reservoir. After a long time, the ensemble reaches equilibrium $q_{ij}^*$. From the SSL perspective, the goal is to find a suitable distribution $q_{ij}^*$ satisfying some properties. This self-labelling mechanism plays a major role in some modern SSL algorithms \cite{Asano2020,Swav,Seer}, as discussed in the applications.

Here, we treat the isochoric heating problem (for long times) as a maximum entropy (MaxEnt) situation. Let the entropy of the composite system $q_{ij}$ be defined as
\begin{equation}
\label{entropy}
S[q]=-\sum_{ij} q_{ij}\log q_{ij},
\end{equation}
and the normalization $\sum_{ij}q_{ij}=1$ and internal energy (\ref{internalenergy}),
\begin{equation}
\label{internalenergyconstraint}
U=\sum_{ij}\epsilon_{ij}(\lambda)q_{ij},
\end{equation}
are constraints.
In other to account for general (non-thermal) reservoirs, we consider $M+L$ constraints of the form
\begin{eqnarray}
\label{constraints2}
\sum_{ij}f_{mj}q_{ij}=c_m, \hspace{.5cm} \sum_{ij}g_{li}q_{ij}=d_l,
\end{eqnarray}
where $m=1,...M$ and $l=1,...,L$. Finally, we proceed to maximize (\ref{entropy}) subjected to constraints (\ref{internalenergyconstraint}) and (\ref{constraints2}),
\begin{equation}
\sum_{ij} \delta q_{ij}(\log q_{ij}+c+\beta \epsilon_{ij}+\sum_m a_m f_{mj}+\sum_l b_l g_{li})=0,
\end{equation}
with Lagrange multipliers $(c,\beta,a_m,b_l)$. The solution reads
\begin{equation}
\label{GGE}
    q^*_{ij}=C\exp(-\beta \epsilon_{ij}-\sum_m a_m f_{mj}-\sum_l b_l g_{li}),
\end{equation}
where $C=\sum_{ij}\exp(-\beta \epsilon_{ij}-\sum_m a_m f_{mj}-\sum_l b_l g_{li})$. Note that (\ref{GGE}) is a generalized Gibbs ensemble (GGE), which has the Gibbs ensemble $(a_m=b_l=0)$ as a special case. The extra constraints seem unusual at first glance, but they could be necessary in SSL for $q_{ij}^*$ to be the solution of an optimal transport problem, as also discussed in the applications.

\subsection{Adiabatic process} This process is characterized by a change in the weights $\lambda\rightarrow \lambda'$ without changing the distribution ($q_{ij}$ is fixed). Therefore, there is possibly non zero work (\ref{work}), but heat (\ref{heat}) is zero in the process, $Q=0$. For a small increment of weights, $\lambda'=\lambda + \delta \lambda$, the work (\ref{work}) reads
\begin{equation}
    \label{infwork}
    W= \delta \lambda\sum_{ij} q_{ij}\frac{\partial}{\partial \lambda}\epsilon_{ij}(\lambda),
\end{equation}
in first order in $\delta \lambda$. Keep in mind that $\delta \lambda \in \mathbb{R}^W$, so $(\delta \lambda) \partial/\partial \lambda := \sum_w \delta \lambda_w (\partial/\partial \lambda_w)$ is a shorthand notation.

Physically, using the same analogy with the gas, one could think of a thermally insulated gas subjected to a volume expansion. Another useful quantity is the irreversible work. For the increment $\delta \lambda$, it reads in adimensional units,
\begin{equation}
\label{workirr}
    W_{irr}:=\beta W-(\delta \lambda)\frac{\partial}{\partial \lambda}F(\lambda),
\end{equation}
with work from (\ref{infwork}) and $F$ is the free energy defined as $F=\sum_i F_i(\beta,\lambda):=-\sum_i \log Z_i(\beta,\lambda)$ and $Z_i(\beta,\lambda):=N\sum_{j}e^{-\beta \epsilon_{ij}(\lambda)}$ is the partition function of the data point $i$. The irreversible work measures gap between the actual work and the work obtained in the reversible (equilibrium) process. This quantity is often depicted in fluctuation theorems as a measure of irreversibility \cite{Jar1997}.

\subsection{Swap process}
Consider a combined system formed by two independent systems $(I_i,y_j)$ and $(\tilde{I}_i,\tilde{y}_j)$ with the same dimensions, $i=1,...,N$ and $j=1,...,K$. The occupations are $q_{ij}$ and $\tilde{q}_{ij}$ respectively. The energy matrix is possibly different, $\epsilon_{ij}(\lambda)$ and $\tilde{\epsilon_{ij}}(\lambda)$, because the data points $I_i$ and $\tilde{I}_i$ are different, although the function $\phi_\lambda$ could be the same. Therefore the combined system has energy $\langle E \rangle = \sum_{ij} q_{ij}\epsilon_{ij}(\lambda)+\tilde{q}_{ij}\tilde{\epsilon}_{ij}(\lambda)$. 
A swap operation, as the name implies, swaps the occupations $q_{ij}$ and $\tilde{q}_{ij}$. This operation produces a final energy 
\begin{equation}
\label{swapenergy}
    \langle E \rangle = \sum_{ij} \tilde{q}_{ij}\epsilon_{ij}(\lambda)+q_{ij}\tilde{\epsilon}_{ij}(\lambda),
\end{equation}
and the resulting energy variation is given by the heat (\ref{heat}). 
\begin{equation}
Q=\sum_{ij}(q_{ij}-\tilde{q}_{ij})(\epsilon_{ij}(\lambda)-\tilde{\epsilon}_{ij}(\lambda)).
\end{equation}
This process is useful in the construction of twin architectures, where are a pair of distorted views ($I_i$ and $\tilde{I}_i$) are passed to the network simultaneously. 

Other processes are also possible. For instance, to push the analogy with thermodynamics even further, a isothermal process would change the weights $\lambda\rightarrow \lambda'$ (work is generated) and heat exchange is also allowed, $q\rightarrow q'$, but in a way that keeps the internal energy (\ref{internalenergyconstraint}) fixed. Using this idea, a Carnot half cycle could also be defined.

\section{Learning Demon}
Now that we introduced some thermodynamic processes in the last section, we frame a general learning problem
as two step cycle, resembling a Otto engine half cycle, as depicted in FIG. 1. For each iteration $t$, the system is (I) subjected to a isochoric heating for large times, resulting in the ensemble $q_{ij}^*$ (\ref{GGE}) for a given $\lambda^t$ as a result of MaxEnt. Then, the system is subjected to (II) a adiabatic transformation for a small $\delta \lambda$, generating work and a final weight $\lambda^{t+1}=\lambda^t + \delta \lambda^t$, when the cycle repeats. 

The demon's challenge is to choose $\delta \lambda^t$ wisely so it extracts negative (irreversible) work (\ref{workirr}), $W_{irr}^t\leq 0$, in each step $t$ of the cycle,
\begin{equation}
\label{wirr}
    W_{irr}^t=(\delta \lambda)\sum_{ij} q_{ij}\frac{\partial}{\partial \lambda}(\beta\epsilon_{ij}(\lambda)-F(\lambda)) \leq 0,
\end{equation}
For the task at hand, the demon is allowed to use knowledge of the system. Namely, the tensor $(\partial/\partial \lambda_w)[\beta \epsilon_{ij}(\lambda)-F(\beta,\lambda)]$ is accessible by the demon in each cycle. It means that the simple choice:
\begin{equation}
\label{protocol}
    \delta \lambda_w = -\gamma \sum_{ij} q_{ij}\frac{\partial}{\partial \lambda_w}(\beta\epsilon_{ij}(\lambda)-F(\lambda)),
\end{equation}
for $\gamma > 0 $ (learning rate), results in a negative irreversible work combining (\ref{wirr}) and (\ref{protocol})
\begin{equation}
\label{wirr2}
    W_{irr}^t=-\gamma [\sum_{wij} q_{ij}\frac{\partial}{\partial \lambda_w}(\beta\epsilon_{ij}(\lambda)-F(\lambda))]^2 \leq 0,
\end{equation}
thus solving the demon's challenge. This learning scheme is the usual stochastic gradient descent (SGD), where the algorithm iteratively minimizes a loss function \cite{DeepReview}.

\section{Applications}
In this section, we examine different SSL frameworks with the tools developed in the previous section. For simplicity, we organized the frameworks in three different classes, each one displaying a completely different ensemble $q_{ij}^*$. In all cases, there is work extraction as in FIG. 1., but they differ in the type of ensemble. The major application is the Sinkhorn-Knopp class, which better represents the thermodynamic interpretation as it combines all processes (heating, adiabatic and swap). For completeness, we also mention the degenerate class, which contains ensembles of the type $q_{ij}^D=\delta_{jk}/N$, such as the supervised algorithms and deterministic self-labelling. Finally, we show how some SSL algorithms without pseudo-labeling might be seen as an adiabatic class with a fixed maximum entropy ensemble (uniform), $q_{ij}^U=1/(NK)$.

\subsection{Sinkhorn-Knopp class} 
This class of SSL models uses fast variants of Sinkhorn-Knopp algorithm to solve an optimal transport problem and find an optimal ensemble $q_{ij}^*$ for the self-labelling. The idea was introduced in \cite{Asano2020}, followed by SwaV \cite{Swav} and SEER \cite{Seer}.
From a thermodynamics perspective, the optimal transport is actually seen as a particular heating problem with specific constraints as discussed below.

First, the energy function (\ref{maths}) has a linear form \cite{Asano2020},
\begin{equation}
    \epsilon_{ij}(\lambda)=-\sum_k H_{jk}x_{ki},
\end{equation}
where $x_{ki}=\phi_\lambda(I_i)_k$ is the $k^{th}$ entry of the feature vector $\phi_\lambda(I_i)$ and $H_{jk}$ is matrix of prototypes vectors. It means that the outcome of the deep neural network is composed with a linear projection head, $H\cdot \phi_\lambda(I_i)$, resulting in a $K$ dimensional vector. In this case, the internal energy (\ref{internalenergy}) reads
\begin{equation}
    \langle E_\lambda(x_i,y_j) \rangle = -\sum_{ij}\epsilon_{ij}(\lambda)q_{ij}=-\sum_{ijk} q_{ij}H_{jk}x_{ki}.
\end{equation}
For the self-labelling, the algorithm finds a solution of the following (smooth) optimal transport problem \cite{Cuturi2013}:
\begin{equation}
\label{opt}
    \min_{q\in U} -\langle q\cdot \ln q^G_{1} \rangle + (1/\beta)KL(q||q^U),
\end{equation}
where $\langle a \cdot b \rangle$ is the Frobenius dot product, $q^G_{1}$ is the Gibbs ensemble for $\beta=1$, $U:=\{q\in \mathbb{R}^{N\times K}|\sum_i q_{ij}=1/K, \sum_j q_{ij}=1/N\}\}$ is a transportation polytope \cite{Cuturi2013}, $KL(p||q)=\sum p_{ij} \ln p_{ij}/q_{ij}$ is the Kullback-Leibler (KL) divergence and $q_{ij}^U=1/(NK)$ is the uniform ensemble.  The solution $q_{ij}^*$ of (\ref{opt}) is computationally implemented with a fast version of the Sinkhorn-Knopp algorithm \cite{Cuturi2013,Asano2020}, which gives the name of the class. 

Now we show that the solution (\ref{opt}) is a member of the GGE (\ref{GGE}), therefore drawing the analogy between the optimization problem and the thermodynamic heating process. First, multiplying (\ref{opt}) by $-\beta$ (now a maximization) and using the definition of KL, $q^G$ and (\ref{entropy}) yields
\begin{equation}
\label{opt2}
   \max_{q\in U} S[q] - \beta \langle \epsilon_{ij}q\rangle+\beta\langle q\cdot \ln q_1^G \rangle + \langle q \log q^U\rangle.
\end{equation}
Then, we translate the transportation polytope $U$ using our formalism (\ref{constraints2}), for $f_{mj}=\delta_{mj}$ and $g_{li}=\delta_{li}$, with $c_m=1/K$ and $d_l = 1/N$, resulting in
\begin{equation}
\label{constraintsSwav1}
\sum_{i}q_{ij}=1/K, \hspace{.5cm} \sum_{j}q_{ij}=1/N.
\end{equation}
One could see these particular constraints as spreading the ensemble such that the marginal distributions of data points and labels are uniform. Finally, replacing $q_1^G$ and introducing the underlying Lagrange multipliers for (\ref{constraintsSwav1}) in (\ref{opt2}), the optimization now reads
\begin{equation}
\label{opt3}
      \max_{q} S[q]-\sum_{ij} q_{ij}( \beta\epsilon_{ij} + a_i + b_j + c),
      \end{equation}
where the terms $\sum_{ij} q_{ij} \log q^U_{ij}$ and $\sum_{ij}\ln Z_i q_{ij}$ were absorbed in the constraints. Note that (\ref{opt3}) is a MaxEnt problem with the following solution from the GGE (\ref{GGE}):
\begin{equation}
    \label{GGESwav}
    q_{ij}^*=C\exp(-\beta \epsilon_{ij} -a_j - b_i),
\end{equation}
where $a_j$ and $b_i$ are such that constraints (\ref{constraintsSwav1}) are satisfied and $C$ is a normalization constant. Because of the equivalence of (\ref{opt}) and (\ref{opt3}), we interpret the self-labelling of the Sinkhorn-Knop class \cite{Asano2020,Swav,Seer} as a particular heating problem.

The second step is adiabatic, where a demon tries to to minimize the irreversible work as presented in (\ref{protocol}). For this step, the authors in \cite{Asano2020} already interpreted it as a maximization of mutual information (between representations and self-labels), as a direct application of Gibbs inequality. Actually, in the thermodynamic picture, learning is achieved by extracting negative irreversible work (\ref{wirr2})
\begin{equation}
\label{wirr3}
    W_{irr}^t=\delta \sum_{ij} - q_{ij}\log q_{ij}^G=-\delta \langle \log q_{ij}^G \rangle \leq 0,
\end{equation}
where the average is over $q_{ij}$ and $q_{ij}^G=\exp(-\beta \epsilon_{ij}(\lambda))/Z_i(\lambda, \beta)$ is the Gibbs ensemble. In other words, the demon changes the weights $\lambda$ so that the Gibbs ensemble $q_{ij}^G$ tries to mimic the current ensemble $q_{ij}^*$ obtained in the heating process. Keep in mind that the matrix $H$ is also adjusted $(H_{kj}\rightarrow H_{kj}')$ (learned) in the adiabatic process, as a part of the weight matrix, using the same reasoning as (\ref{protocol}).

The SwaV model \cite{Swav} also generates different views of the initial data point $I_i$. In the case of two views ($I_i, \tilde{I}_i$), a swap operation is performed in the combined system for each cycle, so that the final energy after the swap is given by (\ref{swapenergy}).

\subsection{Degenerate class}
We consider a degenerate ensemble, $q_{ij}^D=\delta_{jk}/N$, meaning that only one class $k$ is assigned for each data point $i$. This is the case of all supervised learning algorithms, but also the case of recent SSL applications that assigns self-labels online using some deterministic clustering technique, such as DeepCluster \cite{Deepcluster}. In this case, the are two alternating steps. The first step is to draw a degenerate ensemble from the representations, $q_{ij}^D=\delta_{jk}/N$, using $k$-means clustering algorithm. One could think of this step as a cooling process at $T=0$, where the composite system has an energy
\begin{equation}
\label{kmeans}
    \epsilon_{ij}^I=||x_i-C_j||^2,
\end{equation}
for $C_j\in\{C_1,...,C_K\}$ vectors of centroid matrix. A cooling process takes the composite system to the Gibbs ensemble $q_{ij}^D=\delta_{ik}/N$ for the limit $T\rightarrow0$, such that $k$ is given by $\min_j \epsilon_{ij}$. For completeness of the $k$-means, the matrix $C$ is also optimized to minimize $(\ref{kmeans})$. 
Then, in the adiabatic step, the algorithm considers some energy $\epsilon_{ij}^{II}$,
\begin{equation}
    \langle E_\lambda(x,y) \rangle = \sum_{ij}\epsilon_{ij}^{II}q_{ij}^D=\frac{1}{N}\sum_{i}\epsilon_{ik}(\lambda),
\end{equation}
which is the general form of the loss function of a supervised learning and some SSL such as DeepCluster. The minimization of the loss function is equivalent to extracting negative work (\ref{infwork})
\begin{equation}
\label{negwork}
    W= \delta \lambda\frac{1}{N}\sum_{i} \frac{\partial}{\partial \lambda}\epsilon_{ik}(\lambda) \leq 0.
\end{equation}
Each SSL cycle alternates between the clustering assignment, with a new degenerate ensemble $q_{ij}^D$ and the extraction of negative work (\ref{work}) in the following step. In summary, one could think of the degenerate class as a zero temperature limit ($T=0$), because the ensemble is deterministic (degenerate) for each data point $i$ (and each cycle, as DeepCluster). The uncertainty in the non deterministic self-labelling is somehow connected to a positive temperature as in the Sinkhorn-Knopp class.

\subsection{Adiabatic class}
In this class, the ensemble $q_{ij}$ is fixed, having the same value for all steps, $q_{ij}^t=q_{ij}^0$ . For that reason, heat exchange (\ref{heat}) is always zero over the cycles, therefore the name of the class. Having a fixed ensemble allows it to be mapped to the maximum entropy ensemble (uniform), $q_{ij}=q_{ij}^U:=1/(NK)$ upon a redefinition of the stochastic energy,  $\epsilon_{ij}(\lambda)q_{ij}^0 \rightarrow \epsilon_{ij}(\lambda)$, resulting in a uniform distribution. Therefore, members of this class will have a energy of the form
\begin{equation}
\label{adiabatic}
    \langle E_\lambda(x,y) \rangle = \frac{1}{NK}\sum_{ij}\epsilon_{ij}(\lambda).
\end{equation}
Intuitively, one could think the adiabatic class as the limit of infinite temperature ($T\rightarrow \infty$), spreading the distribution $q_{ij}$ uniformly over all possible self-labels $k$. For that reason, the pseudo-labelling (heating) step is absent in this class, as the only relevant information is the pseudo-label dimension $K$. This is the case, for instance, of BYOL \cite{Byol} and Barlow-Twins \cite{Barlowtwins}. Although the algorithms do not need to explicitly mention the existence of label classes $K$, we could think of them as the vector indexes of the (final) representation as the underlying classes with a suitable (uniform) ensemble $q_{ij}$. For instance, BYOL has an energy given by
\begin{equation}
\label{BYOL}
    \langle E \rangle = 2-2\sum_{ij}\overline{p}(x_{i}(\lambda))_j \overline{x}_{ij}'(\xi),
\end{equation}
for $\overline{x}_i:=x_i/|x_i|$, with $x_i=\phi_\lambda(I_i)$ the representation generated by the main network and $x_i'=\phi_{\xi} (I_i')$ a representation of a distorted data point generated by a second network. The function $p$ is a normalized prediction, $\overline{p}(z_i)=p(z_i)/|z_i|$, for some function $p$. Comparing (\ref{adiabatic}) and (\ref{BYOL}), one could assign $\epsilon_{ij}=2-2(NK)\overline{p}(x_{i}(\lambda))_j\overline{x}_{ij}'(\xi)$, as representative of the adiabatic class. The final form also includes a swapped term (where $I_i$ and $I_i'$ are exchanged), but the idea remains the same. As usual, learning by minimization of the loss function is equivalent to a negative work extraction, such as (\ref{negwork}).
Other methods, such as Barlow-Twins (BT), follow the same idea as in (\ref{BYOL}), but increasing the sophistication of the stochastic energy.

\section{Summary and conclusions}
We proposed a thermodynamic interpretation for SSL algorithms as a composed system formed by representations and self-labels operating in cycles. After defining stochastic and internal energy, heat and work, we showed that learning is seen as a demon that extract negative (irreversible) work in the adiabatic step using knowledge from the system. Particularly, the heating process is in close analogy to self-labelling with uncertainty, a situation where the optimal transport problem posed by some SSL algorithms (Sinkhorn-Knopp class) translates to a MaxEnt problem, and the solution is a member of the GGE. We also discussed the degenerate and adiabatic classes as the limits $T=0$ and $T\rightarrow \infty$.

In this picture, it is natural to ask if there are other online self-labelling heating processes suitable for SSL cycles beyond the Sinkhorn-Knopp class. In other words, are there other particular GGE classes useful for SSL? For instance, one could change some of the constraints of optimal transport (\ref{constraintsSwav1}) or think of a Carnot half cycle (instead of Otto's), using an isothermal process (with heat and work) instead of a isochoric heating. 

We expect that the interpretation presented in this paper will help the design and understanding of novel SSL algorithms from a thermodynamics standpoint.

\label{conclude}

\end{document}